\documentstyle[twoside,fleqn,espcrc2]{article}

\newcommand{\beq}{\begin{equation}}
\newcommand{\eeq}{\end{equation}}
\newcommand{\bea}{\begin{eqnarray}}
\newcommand{\eea}{\end{eqnarray}}
\newcommand{\Slash}[1]{ \!\not{\!\!#1} }

\newcommand{\AmS}{{\protect\the\textfont2
  A\kern-.1667em\lower.5ex\hbox{M}\kern-.125emS}}

\title{Improved actions of the staggered fermion }

\author{Yubing Luo\address{Columbia University, Department of
	Physics, New York, NY 10027}
    }
       
\begin{document}

\begin{abstract}

We have studied $O(a^2)$ improved lattice QCD with the staggered
fermion by using Symanzik's program. We find that there are 5
dimension-6 fermion bilinears and gauge operators. In addition,
there are 10 four-fermion operators which are absent at the tree-level
and tadpole-improved tree-level.

\end{abstract}

\maketitle

\section{INTRODUCTION}

To remove $O(a^2)$ errors by using Symanzik's program, we must
find all possible dimension-6 terms which are scalars under the
lattice symmetry transformations. Then, we add them to the original
action and adjust their coefficients so that low-momentum physical
quantities do not have $O(a^2)$ errors.

When he proposed this program, Symanzik applied it to scalar
fields and proved the consistency of the improvement program to
all orders of perturbation theory \cite{symanzik}.
L\"uscher and Weisz applied this program to pure gauge theory
and developed the concept of on-shell improved action which demands
only the improvement of all physical (on-shell) quantities
\cite{Luscher}. Sheikholeslami and Wohlert applied
this program to Wilson fermions and proposed a ``Clover'' term
to remove all $O(a)$ errors from Wilson fermions \cite{SW}.
Naik applied this program to Dirac-K\"ahler fermions and
proposed a three-link term \cite{Naik}. Although the staggered fermion
and the Dirac-K\"ahler fermion formalisms are the
same in the free case, they are quite
different when the gauge interaction is included.
Dirac-K\"ahler fermions have two drawbacks: First, the $U(1)_o \otimes
U(1)_e$ symmetry does not prevent the gauge interaction from
producing non-zero mass counterterms, therefore the bare mass
parameters have to be tuned in the continuum limit, similar to
Wilson fermions; Second, there is no massless Goldstone boson
\cite{Napoly}. Therefore, in most applications, the standard
staggered fermion formalism is used. Thus, it is important to
investigate the improved action for the staggered fermion case.

Sharpe's pioneering work showed that there is no $O(a)$ term for the
staggered fermion \cite{Sharpe}. Thus the leading order terms in
the improved action for staggered fermions is $O(a^2)$.
In this talk, I will present all independent dimension-6 terms,
and show which terms remain after a spectrum-conserving
transformation of the fields.
A detailed discussion of these results can be found in
ref.~\cite{Luo}.

\section{IMPROVED ACTION}


The improved action consists of two parts, the gauge action and
fermion action:
\beq
    S = S_G + S_F.
\eeq
The improved gauge action contains 3 independent dimension-6
operators and can be written as:
\beq
    S_G = \frac{2}{g_0^2}\Bigl[c_0{\cal L}_0 + c_1{\cal L}_1 +
    	c_2{\cal L}_2 + c_3{\cal L}_3\Bigr].
\eeq
where ${\cal L}_0$ denotes the loops consists of 4 links, and
${\cal L}_1$, ${\cal L}_2$, and ${\cal L}_3$ are 3 sets
of loops consisting of 6 links.
The 4 coefficients satisfy a normalization condition:
\beq
    c_0 + 8 c_1 + 8 c_2 + 16 c_3 = 1.
\eeq

The improved fermion action can be written as:
\bea
    S_F = \bar\chi (\Slash{\cal D} + m) \chi
    	+ a^2 \sum_{i=1}^{7} b_i {\cal O}_i \nonumber \\
	 + a^2 \sum_{i=1}^{18} b_i' {\cal F}_i
	 + O(a^3).
\eea
It contains 7 fermion bilinears ${\cal O}_i$ in addition to the
usual Dirac action and 18 four-fermion
operators ${\cal F}_i$.

Therefore, there are 25 possible dimension-6 operators which can be
added to the original lattice action.

\subsection{Fermion bilinears}

The 7 fermion bilinears are the linear combinations of the pattern:
${\cal D}^3$, ${\cal D}^2\Slash{\cal D}$,
$\Slash{\cal D}{\cal D}^2$,
${\cal D}\Slash{\cal D}{\cal D}$,
$\Slash{\cal D}^3$, $m\Slash{\cal D}^2$ and 
$m{\cal D}^2$. They can be written as:
\bea
    {\cal{O}}_1 = \bar{\chi}\,{\cal{D}}^3 \chi,		\\
    {\cal{O}}_2 = \bar{\chi}\,\frac{1}{2} \left(
    	{\cal{D}}^2\Slash{\cal{D}}
	-\Slash{\cal{D}}{\cal{D}}^2 \right)\chi,	\\
    {\cal{O}}_3 = \bar{\chi}\, \frac{1}{2} \left(
    	{\cal{D}}^2\Slash{\cal{D}} +
	\Slash{\cal{D}}{\cal{D}}^2
	- 2 \Slash{\cal{D}}^3 \right)\chi,		\\
    {\cal{O}}_4 = \bar{\chi}\, \left(
    	{\cal{D}}^2\Slash{\cal{D}} +
	\Slash{\cal{D}}{\cal{D}}^2
	- 2 {\cal{D}}\Slash{\cal{D}}{\cal{D}}
	\right)\chi,						\\
    {\cal{O}}_5 = \bar{\chi}\,\Slash{\cal{D}}^3 \chi.	\\
    {\cal{O}}_6 = m\,\bar{\chi}\,\Slash{\cal{D}}^2 \chi,\\
    {\cal{O}}_7 = m\,\bar{\chi}\,{\cal{D}}^2 \chi.
\eea
${\cal{O}}_1$ is the Naik's term. ${\cal{O}}_4$ is equivalent to the
MILC's ``fat link'' term~\cite{MILC}. ${\cal{O}}_6$ and ${\cal{O}}_7$
violate the $U_A(1)$ symmetry and therefore depend
explicitly on the mass parameter.

\subsection{Four-fermion operators}

The 18 four-fermion operators can be divided into 4 groups. The
first group contains one operator:
\beq
    {\cal F}_1 = \sum_{x,a} \bar\chi(x) t^a\chi(x)
    \sum_e \bar\chi(x\!+\!e)t^a\chi(x\!+\!e),
\eeq
where $t^a$ are the 8 generators of $SU(3)$ color group
and the sum over $e$ is a sum over the 8 possible lattice
displacements of length ``1''.
The second group contains one operator:
\beq	{\label{chi_form:2}}
    {\cal{F}}_2 = \sum_{x,a} \bar\chi(x) t^a\chi(x)
    	\sum_v \bar\chi(x\!+\!v)t^a\chi(x\!+\!v),
\eeq
where the sum over $v$ is over the 32 possible lattice displacements
of length ``$\sqrt{3}$~''.
The third group contains 8 operators:
\bea
    {\cal{F}}_i = \sum_{x,a} \sum_\mu
    	{\cal C}^a_\mu(x) \frac{1}{256}
	\sum_c w(c)\eta_5(c)	\nonumber \\
	P^{(i)}_\mu(c) {\cal C}^a_\mu(x+c)
		\\
    i = 3, \cdots , 10.		\nonumber
\eea
where the sum over $c$ is a sum over the 81 displacements with
coordinates $c_\mu = -1, 0, 1$. The weight is:
\beq
    w(c) = \prod_{\mu=1}^{4}\,(2-\left|c_\mu\right|).
\eeq
The fermion bilinear ${\cal C}^a_\mu(x)$ is given by:
\beq
    {\cal C}^a_\mu(x) = \bar\chi(x) t^a \sum_{v\perp\hat\mu}
	\chi(x\!+\!v),
\eeq
where the sum is over the 8 possible lattice displacements of length
``$\sqrt{3}$~'' which are perpendicular to $\hat\mu$ direction.
The last group also contains 8 operators:
\bea {\label{four-fermi:4}}
    {\cal{F}}_i = \sum_{x,a} \sum_{\mu} {\cal B}^a_\mu(x)
	\frac{1}{256}\sum_c w(c)	\nonumber \\
	P^{(i)}_{\mu}(c){\cal B}^a_\mu(x+c),
		\\
    i = 11, \cdots , 18,		\nonumber
\eea
with the fermion bilinear
\beq
    {\cal B}^a_\mu(x) = \frac{1}{2}\Bigl[
        \bar\chi(x) t^a\chi(x\!+\!\hat\mu)
        + \bar\chi(x) t^a\chi(x\!-\!\hat\mu) \Bigr].
\eeq

\subsection{Tree-level values}

At the tree-level, we got:
\bea
    c_0 = \frac{5}{3}, \qquad
    c_1 = -\frac{1}{12}, \qquad
    b_1 = -\frac{1}{6},	\\
    b'_{12} = \frac{g_0^2}{8}, \qquad
    b'_{13} = \frac{g_0^2}{24}, \qquad
    b'_{14} = \frac{g_0^2}{16}.
\eea
All other coefficients vanish at the tree-level.

For staggered fermions, the high-momentum gluon exchange gives rise
to flavor changing four-fermion terms{\footnote{We thank G. P.
Lepage for pointing out the existence of such tree-level
contributions.}. However, these terms belong to a restricted class
of operators which can be expressed as a product of two fermion
bilinears, with each such bilinear composed of fields $\bar\chi(x)$
and $\chi(x')$ with a distance between $x$ and $x'$ of precisely ONE
link. As we will point out in the next section, all the coefficients
of the 8 such terms in Eq.~(\ref{four-fermi:4}) can be changed by a
transformation of the field variables and a choice of fields can be
found for which these coefficients are zero. Thus, the 8 terms in
Eq.~(\ref{four-fermi:4}), including the three tree-level terms
quoted above, actually do not appear in the on-shell improved action.

\subsection{On-shell improved action}

Given one improved action, we can obtain another one by a
transformation of the fields. However, all these actions are
equivalent because all such actions will give the same value for a
specific on-shell quantity. Thus, we can choose to minimize the
number of operators occurring in the on-shell improved action by an
appropriate definition of the field variables.

By redefining both gauge and fermion field variables, we find the
following operators are redundant: ${\cal O}_2$, ${\cal O}_3$, 
${\cal O}_5$, ${\cal O}_6$, ${\cal F}_{11} \cdots {\cal F}_{18}$.
So, we can always find a spectrum conserving transformation of gauge
and fermion fields such that the following coefficients can be set
to be zero:
\bea
    b_2 = b_3 = b_5 = b_6 = 0, \nonumber	\\
    b'_{11} = \cdots = b'_{18} = 0.
\eea
Can we transform away more four-fermion operators? The answer is
``no''. Because in the staggered fermion action, $\bar\chi$ and
$\chi$ are separated by only one link. After the field
transformation, the $\bar\chi$ and $\chi$ in the fermion bilinears
in the resulting four-fermion operators are separated by only one
link also. Hence, only the four-fermion operators in 
Eq.~(\ref{four-fermi:4}) can be redundant.

Under the field transformation, operators ${\cal L}_3$
can give a contribution to ${\cal O}_4$, so we can either
set $c_3 = 0$ or $b_4 = 0$. Thus we have two equivalent improved
actions:
\bea
    S^{I} = c_0{\cal L}_0 + c_1{\cal L}_1 + c_2{\cal L}_2
    		\qquad \qquad \nonumber \\
	+ a^2 \Bigl[b_1{\cal O}_1 + b_4{\cal O}_4 + b_7{\cal O}_7
	+ \sum_{i=1}^{10}b'_i {\cal F}_i\Bigr],
\eea
or
\bea
    S^{II} = c_0{\cal L}_0 + c_1{\cal L}_1 + c_2{\cal L}_2
    	+c_3{\cal L}_3
    		\nonumber \\
	+ a^2 \Bigl[b_1{\cal O}_1 + b_7{\cal O}_7 +
		\sum_{i=1}^{10}b'_i {\cal F}_i\Bigr].
\eea
In either case, the total number of dimension-6 operators is 15.

\section{CONCLUSIONS}

The on-shell improved action for lattice QCD with staggered
fermions contains 15 dimension six operators. Ten of these are
four-fermion operators, which are absent at the tree-level,
and hence of the order of $O(g_0^4a^2)$ at most. The other 5 are
fermion bilinears and gauge operators and only two of them are
nonzero at tree-level. The numerical results from the MILC and
Bielefeld groups presented during this conference are consistent
with our analysis.

\bigskip
\bigskip

I thank Prof. Norman H. Christ for the
extensive discussions during every stage of this work.
This research was supported by the US Department of Energy under
grant DE-FG02-92 ER40699.

\end{document}